\documentclass[11pt,a4paper]{article}

\usepackage{graphics}

\begin{document}

\flushbottom
\itemsep 0pt
\parsep 0pt
\baselineskip 12pt
\parindent 7mm
\vspace*{36pt}

\noindent {\bf A REALISTIC INTERPRETATION OF LATTICE GAUGE \\ THEORIES}

\vskip 36pt 
\hskip 1cm{ \bf Miguel Lorente}

\vskip 12pt

\hskip 1cm {\it Departamento de F\'{\i}sica, Facultad de Ciencias}

\hskip 1cm {\it Universidad de Oviedo}

\hskip 1cm {\it E-33007, Oviedo,  Spain}

\vskip 24pt

\baselineskip 13pt
\noindent
Following recent assumptions to unify quantum mechanics and general relativity, the 
structure of spacetime is suppose to be a consequence of the relations among some 
fundamental objects, and its concept can be formulated without the reference to the intuition. As physical consequences 
the continuous laws should be translated in to difference equations and the lattice 
field theories should be interpreted as a realistic model.

\vskip 13pt


\vskip 26pt

\noindent \parbox[t]{7mm}{\bf 1.}  \parbox[t]{119mm}{\bf RECENT ASSUMPTIONS TO UNIFY
QUANTUM  MECHANICS AND THE STRUCTURE OF SPACETIME}

\vskip 13pt

\indent 
One of the most difficult problems to unify the postulates of QM and general relativity
is  the different conceptions of spacetime. In QM the spacetime is a container where 
the fields are distinguished by their position and interactions, in the theory of 
relativity the gravitational field is identified with the metrical properties of spacetime.
Recently several authors have tried to overcome this difficulty by deriving the  structure of
spacetime from the properties of fundamental processes described by  QM.

According to Joseph M. Jauch,  the set of propositions of a physical system in QM replaces 
the phase space [1]. In classical mechanics the underlying spacetime is necessary to 
impose the initial conditions that determined the solution of equations of motion.
In QM the equation of motions are substituted by the set of propositions based on the 
superposition of the simplest yes-no experiments. The axiomatic form of this 
structure gives rise to the calculus of propositions, that do not presuppose the space
time. The physical state is the result of a series of physical manipulations on the set of 
propositions.

\baselineskip 13pt

Karl F. von Weizsaecker  gives a  more explicit connections between quantum theory and the
concept of time and space  [2]. All the quantum processes can be reduced to binary
alternatives (equivalent to  yes-no experiments). The interaction among these fundamental
entities, which he calls ``urs'', gives rise to  physical system and the structure of
spacetime is the set of relations among the  binary alternatives.

There are two important postulates: (i)  the number of actual alternatives that 
determine a physical is finite, because they represent real properties; (ii)  the number 
of possible alternatives is infinite  due to the indeterministic nature of the quantum 
processes. As a consequence, the description of facts is given by discrete variables, 
but the physical laws are given in terms of continuous functions.

Roger Penrose  does  not pressupose an underlying spacetime for the physical processes [3].
The starting  point is the total angular momentum of some fundamental units, the interactions
of  which produce a discrete network. ``My model, says Penrose, works with objects and 
interactions between objects. An object is thus located either directionally or  positionally
in terms of its relations with other objects. One does not really need a  space to begin with.
The notion of space comes out as a convenience at the end.''

According to David Finkelstein  the world is a network of quantum processes, which he calls 
``monads''[4]. Every process in nature is a finite assembly of elementary processes,  namely,
of creation and destruction, and the structure of Spacetime is the set of all  elementary
processes and their relations.

\vskip 26pt

\noindent \parbox[t]{7mm}{\bf 2.}  \parbox[t]{119mm}{\bf EPISTEMOLOGY OF THESE MODELS}

\vskip 13pt

\indent
In order to understand better these models it would be useful to consider three levels 
of human knowledge in the comprehension of the physical world[5]:

\vskip 13pt

{\noindent {\it Level 1}: Physical magnitudes, such as distances, intervals, force, mass,
charge,  that are given by our sensation and perceptions.}

{\noindent {\it Level 2}:  Mathematical structures, that are the result of metrical properties given 
by measurements and numerical relations among them.}

{\noindent {\it Level 3}: Fundamental concepts, representing the ontological properties of 
physical world given by our intelligence in an attempt to know the reality. This level 
of knowledge is not accepted by some philosophical positions like logical positivismus, 
conventionalismus, neokantismus.}

\vskip 13pt

There must be some connections between the three levels. In QM the theoretical models 
of microphysics in level 2 are related to observable magnitudes in level 1 by correspondence
laws. If we  accept level 3 it should be connected to level 2 and to level 1 (through level
2). In  fact, the rules governing the constructions of theoretical models in level 2 must be 
grounded in some fundamental (ontological) properties of the physical world.

We can now raise the following question: in theoretical models of level 2 there are 
primitive and derived concepts, the last ones are obtained from the first ones by 
mathematical formulas. Are space and time primitive or derived concepts?
If the second answer is given the description of the world in level 2 should not include as 
primitives the geometrical objects such as lines, planes, surfaces.

\vskip 26pt

\noindent \parbox[t]{7mm}{\bf 3.}  \parbox[t]{119mm}{\bf MODERN THEORIES ON THE STRUCTURE OF 
SPACETIME}

\vskip 13pt

\indent 
In order to answer the last question it is convenient to recall the different 
interpretations of the concepts of space and time [6]. They are usually divided in 
three classes.

\begin{enumerate}
\item[(a)]{\it Dualistic theories}:  Space is a container where the particles and waves are 
moving. Time is also a separated entity with respect to which the motion takes place. 
Therefore space and time are absolute and can be thinked of in the absence of 
particles (Newton).
\item[(b)]{\it Monistic theories}:  Spacetime is identified with some properties of matter
and  can not be concevible without the existence of the later. The field of forces and also
the  sources are nothing more that geometrical deformations of the Spacetime (Einstein, 
Kaluza-Klein, Wheeler).
\item[(c)] {\it Relational theories}:  Spacetime consists of the set of relations among some 
fundamental objects: monads (Leibniz), units (Penrose), processes (Weisaecker, Finkelstein),
preparticles (Bunge, Garc\'{\i}a Sucre), objects (Hilbert).
\end{enumerate}

In Sec. 1 we have mentioned some of these authors. We expand in some detail 
Leibniz's and Hilbert's conception. According to Leibniz [7] ``time is the order of points 
(monads) non existing simultaneously and one is the ratio of the other. Space is the 
order of points that exist simultaneously and are connected by mutual interactions. 
Space is nothing more that the set of all points and their relations. One point is here if 
it has relations with some particular points around it. A point changes its position if it 
changes its relations from some points to different ones. Motion is the change of 
different positions in time''.

In his {\it Foundation of Geometry}, Hilbert has proposed an axiomatic approach to 
Euclidean geometry[8], according to which the concept of space is constructed with the 
help of some logical properties. He distinguishes two types of axiomatization: i) 
material, by which the concept of space is taken from observation and intuition and ii) 
formal, in which the concept of space is derived from some formal properties of 
axioms and inferences without the recourse to the intuition or the observation (his 
famous expression, ``We could say always instead of points, lines and planes, chairs, 
tables and glasses of beer,'' confirms his position in favor of the formal axiomatization) 
The concepts of point, straight line, and plane can be reduced to pure logical relations.

\vskip 26pt

\noindent \parbox[t]{7mm}{\bf 4.}  \parbox[t]{119mm}{\bf A RELATIONAL THEORY OF SPACETIME}

\vskip 13pt

\indent
Following the assumption of the last section now we give an explicit construction of a  formal
structure of Spacetime, without the recourse to intuition. We can think of a set of
fundamental objects acting among themselves, giving rise to a  network of relations. These
relations do not pressupose some space. The objects are  nowhere if we consider them as
elements of the physical world in level 2. In order to  be specific we take as a naive network
a three-dimensional cubic lattice. Obviously the network can be taken with different
structure, such as, triangular,  quasiperiodic or random lattices. In order to make connection
with the euclidean geometry we take, for simplicity, a  infinite set of interacting points in
the relation 1 to 4. The set of all relations form a  two-dimensional lattice, in which we can
define:

\vskip 13pt

A {\it path} is the connection between two different points, say, A and B, through 
points that are pairwise neighbours.

The {\it length} of a path is the numbers of points contained in the path, including the 
first and the last one.

A {\it minimal path} is a path with minimal length (in the picture the two paths 
between A and B are minimal). Between two point there can be different minimal 
paths.

\bigskip
\begin{center}
\includegraphics{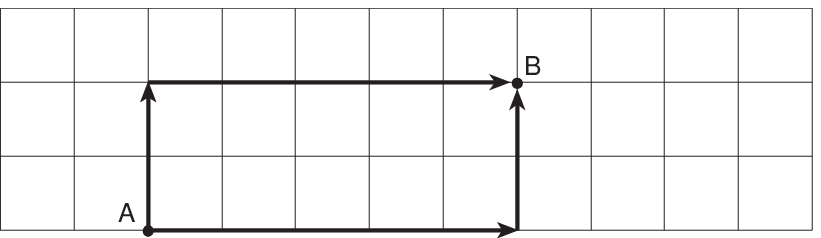}
\end{center}
\medskip

A {\it principal straight line} is a indefinite set of points in the lattice, such that each 
of them is contiguous to other two, and the minimal path between two arbitrary 
points of this line is always unique.

\vskip 13pt

{\noindent {\bf Theorem 1}. Through a point of a 2-dimensional square lattice pass only two 
different principal straight lines (they are called {\it orthogonal straight lines}).}

\vskip 13pt

{\noindent {\bf Theorem 2}. Two principal straight lines that are not orthogonal have all the
points  either in common or separated (in the last case they are called {\it paralell straight 
lines})}.

\vskip 13pt

From these two theorem we can define Cartesian (discrete) coordinates and an 
Euclidean space where the postulates of Hilbert can be applied (with the exception of 
the axioms of continuity). This structure of 2-dimensional space can be easily generalized to
3-dimensional cubic lattice. As we mentioned, those assumptions for the structure of space are
given in level 2, but  it corresponds to the properties of physical space described in level 1
by our  sensations.

In order to introduce the relation that correspond to time we 
start with only two fundamental objects acting among themselves:

\bigskip
\begin{center}
\includegraphics{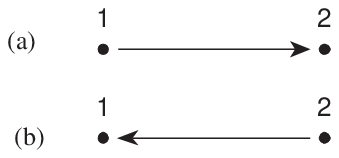}
\end{center}
\medskip
\vskip 13pt

In (a), 1 is acting on 2, and in (b) 2 is acting on 1. But the action of 1 on 2 is supposed 
to be a necessary condition for the action of 2 on 1, and similarly the action of 2 on 1 
is supposed to be a necessary condition for a new action of 1 on 2. Thus we can think 
of a chain of mutual interactions arranged in a series of necessary conditions.
This picture has to be enlarged for the whole lattice. We take a set of interacting 
objects in the relations 1 to 2.

\bigskip
\begin{center}
\includegraphics{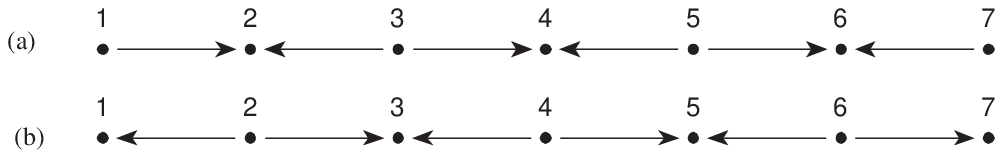}
\end{center}
\medskip

\vskip 13pt

In (a), 1 is acting on 2, 3 is acting on 2 and 4, 5 is acting on 4 and 6, 7 is acting on 6.
In (b), 2 is acting or 1 and 3, 4 is acting on 3 and 5, 6 is acting on 5 an 7.

We postulate that the actions of (a) are necessary conditions for the actions of (b) and 
the actions of (b) are necessary conditions for a further action of type (a) an so on.

Now take a network of objects acting in the relation 1 to 4.

\bigskip
\begin{center}
\includegraphics{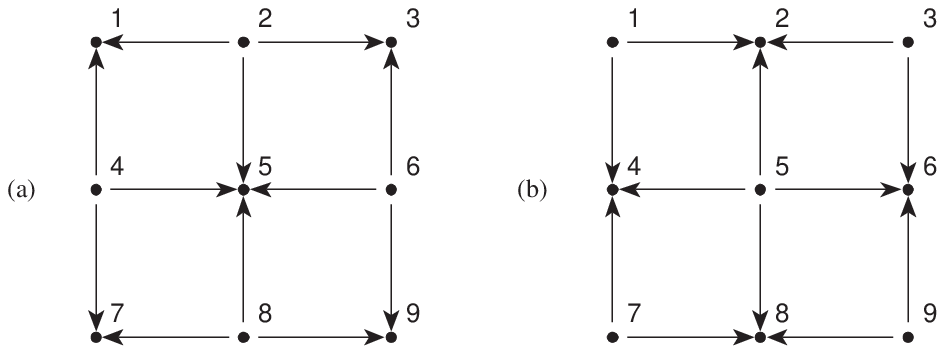}
\end{center}
\medskip

\vskip 13pt

In (a), 2 is acting on 1, 3, 5; 4 is acting on 1, 5, 7; 6 is acting on 3, 5, 9; 8 is acting on 
5,7,9. In (b), 1 is acting on 2 and 4; 3 is acting on 2 and 6; 5 is acting on 2, 4, 6, 8; 7 is
acting  on 4 and 8; 9 is acting on 6 and 8. As before we postulate that the actions of (a) be
necessary conditions for the actions  of (b) and so on. These logical properties of
interactions belong to level 2 and do not pressupose the  concept of time, but they can be put
in correspondence with the physical properties of  time given in level 1.

In level 3 there must be some ontological properties corresponding to the objects and 
interactions described in level 2.

In our model the most essential character of material entities is not the extension but 
their capacity to produce effects in other object (external causality). There is a causal 
relation between cause and effect and the logical necessity that was introduced in the 
last paragragh for the interpretation of time has its ontological ground in the principle 
of causality by wich the effect cannot be produced without its cause.

\vskip 26pt

\noindent \parbox[t]{7mm}{\bf 5.}  \parbox[t]{119mm}{\bf PHYSICAL CONSEQUENCES OF THE MODEL}

\vskip 13pt

\indent
The assumption of relational theory of Spacetime with a particular structure of cubic 
lattice, implies some physical consequences for the classical as well for the quantum 
physics:
\begin{enumerate}
\item[(i)] The Spacetime is discrete, therefore the physical laws are written in the language
of  finite differences. The solutions have to be described by continuous function of 
discrete variables [10].
\item[(ii)] Lattice gauge theories are not only a mathematical tool but a realistic theory, 
because they correspond to the underlying discrete structure of Spacetime. Some 
correspondence law must be given to make connection with the experimental world [11].
\item[(iii)] The symmetry of the model is still Poincar\'e transformation, although one has to 
select those integral transformations that keep the lattice invariant [12].
\item[(iv)] Some experimental test.Although there are infinite number of integral Lorentz
transformations, and the  continuous Lorentz transformations is a dense set, there are only 24
 pure rotations  that keep the lattice invariant. Therefore there is a broken SO (3)
symmetry that  leads to non-isotropy of the world. This means that one could fine in principle
some  preferred direction either in the microphysical world or in the large scale of the 
universe. An other physical application of the model could be detected in the discrete 
mass spectrum as a natural consequence of the elementary time interval an 
estimation of wich by actual calculations gives about $10^{-36}$ sec.
\end{enumerate}

In order to prove this  we summarized some mathematical results of lattice field
theories [13].

We introduce the method of finite differences for the Klein-Gordon scalar
field. An explicit scheme for the wave equation consistent with the continuous
case (the truncation error is of second orden with respect to space and time
variables) can be constructed as follows:
\begin{equation}
\left( {{1 \over {\tau}^{2}}{\nabla}_{n} {\Delta
}_{n} {\tilde{\nabla}}_{j} {\tilde{\Delta}}_{j} -{1 \over
{\varepsilon}^{2}}{\nabla}_{j} {\Delta}_{j}
{\tilde{\nabla}}_{n} {\tilde{\Delta}}_{n} +{M}^{
2}{\tilde{\nabla}}_{n} {\tilde{\Delta}}_{n} {\tilde{\nabla
}}_{j} {\tilde{\Delta}}_{j}}\right) {\phi}_{j}^{n}
=0 \; ;
\label{3.1}
\end{equation}

\noindent
here in the field is defined in the grid points of the $(1+1)$-dimensional
lattice $ \phi_j^n \equiv \phi \left(j\varepsilon,n\tau\right),\;\varepsilon
, \tau $ being the space and time fundamental intervals, $j, n$ integer
numbers, and ${\Delta}_{j} \left({{\nabla}_{j}}\right)$ are the
forward (backward) differences with respect to the space index, ${\tilde{
\Delta}}_{j} \left({{\tilde{\nabla}}_{j}}\right)$ the forward
(backward) averages, and similarly for the time index.

Using the method of separation of variables, it can easily be proved that the
following functions of discrete variables are solutions of the wave equation
(\ref{3.1}):
\begin{equation}
{f}_{j}^{n} \left({k , \omega}\right)
={\left({{1+{1 \over 2} i \varepsilon k \over 1-{1
\over 2} i \varepsilon k}}\right)}^{j}
{\left({{1-{1 \over 2} i \tau \omega \over 1+{1 \over
2} i \tau \omega}}\right)}^{n} \; ,
\label{3.2}
\end{equation}
provided the ``dispersion relation'' is satisfied:
\begin{equation}
{\omega}^{2}-{k}^{2}={M}^{2}
\label{3.3}
\end{equation}
M, being the mass of the particle.

In the limit, $j\rightarrow \infty,\quad n \rightarrow \infty,\quad 
j\varepsilon \rightarrow x,\quad n\tau \rightarrow t $, the functions (\ref{3.2})
become plane wave solutions
\begin{equation}
{f}_{j}^{n} \left({k , \omega}\right) \rightarrow
\exp i\left({kx - \omega t}\right)\; .
\label{3.5}
\end{equation}

Imposing boundary conditions on the space indices,
\begin{equation}
{f}_{o}^{n} \left({k , \omega}\right) ={f}_{
N}^{n} \left({k , \omega}\right)
\label{3.6}
\end{equation}
we get

\begin{equation}
k_m ={2 \over\varepsilon} \tan {\pi m \over N} \; ,\qquad m=0,1,\ldots ,N-1 \; ,
\label{3.7}
\end{equation}
therefore 
\begin{equation}
{\omega =\pm \left({{k}_{m}^{2}+{M}^{2}}\right)}^{1/2}\; .
\label{3.8}
\end{equation}
For the positive energy solutions we define
\begin{equation}
{\omega_m = + \left({{k}_{m}^{2}+{M}^{2}}\right)}^{1/2} \; .
\label{3.9}
\end{equation}

Starting from the wave equation (\ref{3.1}), we can construct a current vector.
Multiplying (\ref{3.1}) by ${\tilde{\nabla}}_{n} {\tilde{\Delta
}}_{n} {\tilde{\nabla}}_{j} {\tilde{\Delta}}_{j}
{\phi}_{j}^{* n}$ from the left, and multiplying the complex
conjugate of the wave equation by ${\tilde{\nabla}}_{n}
{\tilde{\Delta}}_{n} {\tilde{\nabla}}_{j} {\tilde{\Delta
}}_{j} {\phi}_{j}^{n}$ from the right, substracting both results, 
 we obtain the ``conservation law''
\begin{equation}
{1 \over \varepsilon} {\nabla}_{j} {j}_{1}-
i {1 \over \tau} {\nabla}_{n} {j}_{4}=0 \; ,
\label{3.10}
\end{equation}
where

\begin{eqnarray}
j_1&\equiv & i \left[{1 \over \varepsilon} \Delta_j \left(\tilde{\nabla}_n 
\tilde{\Delta}_n \phi_j^{* n}\right) \tilde{\Delta}_j \left(\tilde{\nabla}_n
\tilde{\Delta}_n \phi_j^{n}\right) \right. \nonumber\\
& &- \left. \tilde{\Delta}_j \left(\tilde{\nabla}_n \tilde{\Delta}_n \phi_j^{*n}\right) 
{1\over \varepsilon} \Delta_j \left(\tilde{\nabla}_n \tilde{\Delta}_n
\phi_j^n\right)\right] \; , \label{3.11}\\
j_4&\equiv & i \rho \equiv \left[{1 \over \tau}
{\Delta}_{n} \left(\tilde{\nabla}_{j} \tilde{\Delta}_j \phi_j^{* n}\right)
\tilde{\Delta }_n \left(\tilde{\nabla}_j \tilde{\Delta}_j {\phi}_{j}^{n}\right) \right.
\nonumber\\ 
& &- \left. \tilde{\Delta}_{n}
\left({{\tilde{\nabla}}_{j} {\tilde{\Delta}}_{j} {\phi
}_{j}^{* n}}\right) {1 \over \tau} {\Delta}_{n}
\left({{\tilde{\nabla}}_{j} {\tilde{\Delta}}_{j} {\phi
}_{j}^{n}}\right)\right]
\label{3.12}
\end{eqnarray}
are the spatial and time component, respectively, of the charge vector current
on the lattice.

The charge density $\rho$ suggest that we can substitute the scalar field
$\phi \left({x , t}\right)$ by the smeared field ${\tilde{\nabla}}_{ j}
{\tilde{\Delta}}_{j} {\phi}_{j}^{n}$ and $\phi^*\left({x,t}\right)$ by
${\tilde{\nabla}}_{ j} {\tilde{\Delta}}_{j} {\phi}_{j}^{*n}$.

A suitable Hamiltonian for the real field $\phi_j^n$ and its conjugate momentum $\pi_j^n$ can
be defined as follows:
\begin{eqnarray}
H_n &= &\varepsilon \sum\nolimits\limits_{j =0}^{N
-1} {1 \over 2}\left\{\left(\tilde{\nabla}_j \tilde{\Delta}_j \pi_j^n\right)^2-{1
\over {\varepsilon }^2}\left(\nabla_j \Delta_j \phi_j^n \right)
\left(\tilde{\nabla}_{j} \tilde{\Delta }_{j} {\phi}_{j}^{n}\right) \right. \nonumber \\
& & \left. +{M}^{2}\left(\tilde{\nabla}_{j} \tilde{\Delta}_{j}
{\phi}_{j}^{n}\right)^{2}\right\}\equiv \varepsilon  \sum\nolimits\limits_{j =0}^{N -1}
{\cal H}_{j}^{ n} \; . \label{3.14}
\end{eqnarray}

As in the continuous case, we can derived the Hamilton equations of motions,
varying the Hamiltonian density ${\cal H}_j^n$ first with respect to the
promediate momentum and secondly with respect to scalar field:

\begin{eqnarray}
{1 \over \tau} {\Delta}_{n} \left({{\tilde{\nabla
}}_{j} {\tilde{\Delta}}_{j} {\phi}_{j}^{n}}\right) &=&
{\partial {\cal H}_{j}^{ n} \over \partial \left({{\tilde{\nabla
}}_{j} {\tilde{\Delta}}_{j} {\pi}_{j}^{n}}\right)}
={\tilde{\Delta}}_{n} \left({{\tilde{\nabla}}_{j}
{\tilde{\Delta}}_{j} {\pi}_{j}^{n}}\right) \; , \\
{1 \over \tau} {\Delta}_{n} \left({{\tilde{\nabla
}}_{j} {\tilde{\Delta}}_{j} {\pi}_{j}^{n}}\right) &=&
{}-{\partial {\cal H}_{j}^{ n} \over \partial \left({{\tilde{\nabla
}}_{j} {\tilde{\Delta}}_{j} {\phi}_{j}^{n}}\right)} \nonumber \\
&=& {\tilde{\Delta}}_{n} \left({{1 \over {\varepsilon}^{
2}}{\nabla}_{j} {\Delta}_{j} {\phi}_{j}^{n} -{
M}^{2}{\tilde{\nabla}}_{j} {\tilde{\Delta}}_{j} {
\phi}_{j}^{n}}\right) \; .
\end{eqnarray}

Applying the difference operator ${1 \over \tau} {\nabla}_{
n}$ on both sides of (13) and substituting (14) in the result, we
recover the wave equation (\ref{3.1}).

Using (13) and (14), it can easily be proved that the Hamiltonian
(12) is independent of the time index $n$, namely:
\begin{equation}
{\nabla}_{n} {H}_{n} ={\Delta}_{n} {
H}_{n} =0 \; .
\label{3.17}
\end{equation}

Since the plane wave solutions ${f}_{j}^{n} \left({{k}_{m}
,{\omega}_{m}}\right) \left({m =0,1, \ldots , N
-1}\right)$ form a complete set of orthogonal functions, we
can expand the smeared field and its conjugate momentum as
\begin{eqnarray*}
{\tilde{\nabla}}_{j} {\tilde{\Delta}}_{j} {\phi
}_{j}^{n} &=&{1 \over \sqrt {N\varepsilon}}
\sum\nolimits\limits_{m =- N/ 2}^{N/ 2-1} {1 \over \sqrt
{2{\omega}_{m}}} \left({{a}_{m} {f}_{j}^{n}
\left({{k}_{m} ,{\omega}_{m}}\right) +{a}_{m}^{
*}{f}_{j}^{* n} \left({{k}_{m} ,{\omega
}_{m}}\right)}\right) \; ,  \\
{\tilde{\nabla}}_{j} {\tilde{\Delta}}_{j} {\pi
}_{j}^{n} &=&{- i \over \sqrt {N\varepsilon}}
\sum\nolimits\limits_{m =- N/ 2}^{N/ 2-1} \sqrt {{{
\omega}_{m} \over 2}}\left({{a}_{m} {f}_{j}^{n}
\left({{k}_{m} ,{\omega}_{m}}\right) -{a}_{m}^{
*}{f}_{j}^{* n} \left({{k}_{m} ,{\omega
}_{m}}\right)}\right) \; . 
\end{eqnarray*}

In order to make connection of our scheme with the Einstein-de Broglie
relations $E=\hbar \omega, \quad p=\hbar k$, we take, for the period $T$ and
wavelength $\lambda$ of the discrete plane waves functions (\ref{3.2}) and
(\ref{3.6}),
$$T=N\tau, \qquad \qquad \lambda = N\varepsilon $$
and, for the phase velocity,
$${v}_{p} ={\lambda \over T} ={\varepsilon \over \tau}\; .$$
\indent We have defined the wave number and the angular frequency of the wave functions
as:
$${k}_{m} ={2 \over \varepsilon} \tan{\pi m \over N},
\qquad \qquad {\omega}_{m} ={2 \over \tau} \tan{\pi m \over N}
,\qquad \qquad m =0,1,\ldots , N -1 \; .$$
\indent Substituting the Einstein-de Broglie relations in the relativistic expresion 
$E^2-p^2=M^2$ (we use natural units $\hbar=c=1$), we obtain

$${\omega}_{m}^{2}-{k}_{m}^{2}={\omega}_{m}^{
2}\left({1-{{\tau}^{2} \over {\varepsilon}^{
2}}}\right)={\omega}_{m}^{2}\left({1-{1 \over {v}_{p}^{
2}}}\right)={M}^{2} \; . $$

Since the phase velocity and group velocity satisfy $v_pv_g=1$, we have finally
$${\omega}_{m}^{2}={{M}^{2} \over 1-{v}_{g}^{2}} \; ,$$
hence $M$ has $m$-dependent discrete spectrum.

\end{document}